\documentstyle[preprint,aps]{revtex}
\renewcommand{\baselinestretch}{1.2}
\begin{document}
\title{Geometry of the Hilbert space and the Quantum Zeno Effect}
\author{A. K. Pati and S. V. Lawande}
\address{Theoretical Physics Division, 5th Floor, Central Complex,}
\address{Bhabha Atomic Research Centre, Mumbai - 400 085, INDIA.}
\date{\today}
\maketitle

\begin{abstract}
We  show  that  the  quadratic short time behaviour of transition
probability  is a natural consequence of the inner product of the
Hilbert space of  the  quantum  system.  We  provide  a  relation
between  the  survival  probability  and the underlying geometric
structure  such  as  the  Fubini-Study  metric  defined  on   the
projective Hilbert space of the quantum system. This predicts the
quantum  Zeno effect even for systems described by non-linear and
non-unitary evolution equations, within the collapse mechanism of
the wavefunction during measurement process.

\end{abstract}

\vskip 3cm

PACS NO: 03.65.Bz

\vskip 4cm
email: krsrini@magnum.barc.ernet.in

\newpage
\par
Using collapse postulate of quantum measurement theory Mishra and
Sudarshan  \cite{1}  have shown that the life-time of an unstable
system can be prolonged by performing ``frequent  observations''.
In  the  limit  of  infinitely frequent observations the decay is
inhibited, which is referred  to  as  the  quantum  Zeno  paradox
(QZP).  Similarly,  for  stable  systems one can prove that if we
prepare  the  system  in  any  one  of  the  eigenstates  of  the
observable  and  interrupt the unitary evolution of the system in
succession, the transition to other eigenstate is prohibited. The
inhibition of transition for  coherently  evolving  system  under
continuous  observation  is  often  termed as quantum Zeno effect
(QZE) \cite{2}. These predictions  are  based  on  principles  of
linear  quantum  theory  and  on the projection postulates of von
Neumann \cite{3} which results in short  time  behaviour  of  the
transition  probability  that  grows  quadratically with time. In
general, the state of a system  evolves  unitarily  according  to
Schr{\"o}dinger   equation   and  to  obtain  a  result  on  some
observable we need to do a measurement. During  measurement  some
coupling  occurs  between  the system and the measuring apparatus
and the apparatus reads one of the eigenvalues of the  observable
of   the   system.  The  state  of  the  system  which  is  in  a
superposition of different eigenstates corresponding to different
eigenvalues of  the  observable,  collapses  to  that  eigenstate
corresponding  to the eigenvalue being measured. This constitutes
a precise and instantaneous measurement.

\par
Although  the  original  formulation  of  QZP has not been tested
experimentally, the QZE for a stable system is  claimed  to  have
been  observed by Itano et al \cite{4}. Following Cook's \cite{5}
proposal Itano's group made use of a two-level atom coupled to  a
third  level  via optical pulses so as to monitor the atom in its
ground state. Two explain their observations  they  use  the  two
ingredients,  namely,  (i)  the  Schr{\"o}dinger  time  evolution
between  two  successive  measurements  and  (ii)  the  reduction
mechanism  and  their  results were found to be in full agreement
with the theoretical predictions. In a sense this  interpretation
was considered to provide an experimental support of the collapse
postulate and QZE. Subsequently, it was argued by many physicists
\cite{6,7}  that the experiment of Itano et al \cite{4} in no way
verifies von Neumann's collapse  postulate.  It  was  also  shown
\cite{6,7}  that  the  results  of  Itano et al \cite{4} could be
explained by making use of unitary dynamical evolution alone  and
the collapse postulate was, in fact, not necessary.

\par
Recently,  there  have been some arguments to clarify the meaning
of continuous observation within the conventional quantum theory.
The continuous observation or measurement is  understood  as  the
limit  of  a  sequence  of  discrete, instantaneous ones when the
interval between two successive measurements tends to zero (which
amounts to taking N, the number  of  measurements  to  infinity).
Sometimes  back  Ghiradi  et  al  \cite{8}  have  argued that the
complete suppresion of decay of  an  unstable  system  cannot  be
realised  due to time-energy uncertainty relation in the original
context of QZP. Nakazato et al \cite{9} have argued that infinite
limit of frequent measurement cannot be taken due  to  losses  in
the  real  experimental  setups  and  also  due to spreads in the
position and momentum of a quantum  particle.  Also  an  approach
based   on   decoherence  \cite{10}  indicates  that  the  finite
decoherence time  sets  a  limit  on  the  interval  between  two
successive  measurements.  Using  the  geometric ideas of quantum
state space and the generalised uncertainty relation one  of  the
present  authors \cite{11} has shown that there is a limit on the
frequency  of  measurement  in  quantum  Zeno  setups   and   the
previously discussed limits are special cases of this generalised
limit.

\par
In  past  several  years,  there  have  been different approaches
suggested to replace von Neumann's collapse postulate  \cite{12}.
Most   interestingly,   the  QZE  has  been  understood  using  a
continuous measurement model based  on  restricted  Feynman  path
integrals  \cite{13}.  This  results  in an effective Hamiltonian
description that is non-hermitian in nature. The  imaginary  part
of  the  Hamiltonian  takes  into  account the effect of meter on
measured system.  There  has  been  a  stochastic  simulation  of
quantum trajectory approach to understand the QZE for ensemble of
ions \cite{14}. Here, the inhibition of transition is achieved by
destroying  the coherence. Since repeated measurements on quantum
systems have considerable interest in  a  variety  of  situations
\cite{15},  it  would  be  important  to  relax some of the basic
postulates of quantum theory so that the quantum Zeno effect  may
be  realised in wide class of systems. Recently, a nice review of
theoretical and experimental status of the  quantum  Zeno  effect
has been given by Home and Whitaker \cite{hw}.

\par
In  this  letter  we examine the necessity of the Schr{\"o}dinger
quantum evolution between two successive measurements. We  recall
that  the usual proof of QZE, is based on two basic postulates of
quantum  theory,  viz,  (i)  the  state  of  the  system  evolves
according  to Schr{\"o}dinger equation between $[0, \tau], [\tau,
2\tau],......[(N-1)\tau, N\tau]$ and (ii)  the  measurements  are
performed    instantaneously    at    discrete    times    $\tau,
2\tau,.....N\tau$, such that total duration of the process  $T  =
N\tau$  is  held fixed. As pointed out above the necessity of the
collapse postulate has been questioned and it has been shown that
a pure  dynamical  equation  could  predict  the  QZE.  Here,  we
question  the  requirement  of  the  first postulate, namely, the
linear unitary time evolution of the quantum system  between  two
successive measurements. We argue that {\it it is not a necessary
but  only  a  sufficient  requirement for predicting QZE.} All we
need is that the soultion of the evolution equation should belong
to a Hilbert space. We argue that the inner  product  defined  on
the Hilbert gives us the quadratic short-time behaviour, which is
necessary   for   the  occurence  of  QZE,  within  the  collapse
postulate. It is well known from the study of geometry of quantum
evolution \cite{pv,16,17,18} that the inner product also  induces
the  metric  structure  on  the  projective  Hilbert space of the
quantum system. We show that the survival probability is directly
related to this geometric structure  of  the  Hilbert  space.  To
define  the  metric  structure the state vector need not follow a
Schr{\"o}dinger time evolution. Further, we provide  a  criterion
for  observing  the QZE for a finite number of measurements as an
infinite number of frequent measurements is not  possible  within
the  von  Neumann type of measurement scheme. As a consequence of
the geometry of the  Hilbert  space,  one  can  observe  QZE  for
systems   described   by   non-linear   Schr{\"o}dinger  equation
\cite{18} and Gisin's equation \cite{19}.  These  equations  give
rise   to   a   quadratic  short-time  behaviour  for  transition
probabilities. The prediction of  QZE  for  systems  governed  by
non-linear   equations  is  rather  interesting  because  in  the
quadratic time-dependence  regime  the  behaviour  of  transition
probablity  can  depend  on  the  non-linearity  parameter of the
dynamical system. This  may  accelerate  the  inhibition  process
under  repeated  observation  as compared to that expected from a
linear dynamical evolution process.

\par
We  consider  a  quantum system whose state vector $|\Psi(t)> \in
{\cal H}$, where ${\cal H}$ is the Hilbert space  of  the  system
with  dimension  K,  i.e., ${\cal H} = C^K$. We are interested in
measuring some observable $O$ of the system, which is represented
by a Hermitian operator. It has an eigenvalue spectrum  $\{O_n\}$
and a complete set of eigenfunctions $\{|\Psi_n>\}$. The spectrum
is  assumed  to be discrete and non-degenerate. Let the system be
prepared initially in the state $|\Psi(0)>$  which  could  be  an
eigenfunction  of  the observable $O$. Without loss of generality
we can assume that the initial  state  is  normalised  to  unity.
However, at a later time the normalisation of the state vector is
not  guarranted.  Now,  we  allow the system to evolve under some
dynamical law and ask the question: How does the  probability  of
finding  the  system  in the initial state at a later time $\tau$
behaves with $\tau$, where $\tau$ is assumed to be small.  We  do
not  assume the unitary, linear Schr{\"o}dinger quantum evolution
to be satisfied by the state vector $|\Psi(t)>$ in  the  interval
$[0, \tau]$. The survival probability for the system to be in the
initial state is defined as

\begin{equation}
P(\tau) = \biggl|\biggl<{\Psi(0) \over ||\Psi(0)||}|{\Psi(\tau) \over ||\Psi(\tau)||}\biggr>\biggr|^2,
\end{equation}
where  $||\Psi(t)|| = <\Psi(t)|\Psi(t)>^{1\over2}$ denotes the norm of the
state vector defined from the inner product.

\par
We introduce the normalised vector $|\chi(t)> = |{\Psi(t) \over ||\Psi(t)||}>$.
On   Taylor   expanding   $|\chi(\tau)>$   arround
$t = 0$, we obtain

\begin{equation}
|\chi(\tau)> = |\chi(0)> + \tau |{\dot \chi}(0)> +
               {\tau^2   \over   2}|{\ddot \chi}(0)>    +
O(\tau^3).
\end{equation}
Therfore,  the  survival  amplitude  of finding the system in the
initial state for short time is given by
\begin{equation}
<\chi(0)|\chi(\tau)> =  1 + \tau <\chi(0)|{\dot \chi}(0)> +
               {\tau^2   \over   2} <\chi(0)|{\ddot \chi}(0)>    +
O(\tau^3).
\end{equation}
Hence, the  probability  of  finding
the  system  at time $\tau$ in the initial state (which is called
survival probability) is given by
\begin{equation}
P(\tau) =  1 +
         \tau^2 [Re<\chi(0)|{\ddot \chi}(0)>  + (Im<\chi(0)|{\dot \chi}(0)>)^2 ].
\end{equation}
Since  $|\chi(t)>$
preserves the norm $||\chi(t)||$
during   the   time   evolution, one can show that the quantity
$<\chi(t)|{\dot \chi(t)}>$ is purely imaginary and
$Re<\chi(t)|{\ddot   \chi(t)}> = - <{\dot \chi}(t)|{\dot \chi(t)}>$.
Using these facts, we can express
the survival probability as
\begin{eqnarray}
P(\tau) & = & 1 - \tau^2 [<{\dot \chi}(0)|{\dot \chi}(0)>) - (i<\chi(0)|{\dot \chi}(0)>)^2 ] \nonumber\\
        & = & 1 - \tau^2 k,
\end{eqnarray}
where  $k  = [<{\dot \chi}(0)|{\dot \chi}(0)>) - (i<\chi(0)|{\dot
\chi}(0)>)^2 ]$.

The  quantity  $k$  can  also be expressed in terms of the actual
state (unnormalised state vector $|\Psi>$) of the system.
Notice that we can write

\begin{eqnarray}
<{\dot \chi}(0)|{\dot \chi}(0)> = {<{\dot \Psi}(0)|{\dot \Psi}(0)> \over ||\Psi(0)||^2 } + {||{\dot \Psi}(0)|| \over ||\Psi(0)||}^2 - {||{\dot \Psi}(0)|| \over ||\Psi(0)||^3} 2 Re <\Psi(0)|{\dot \Psi}(0)> \nonumber\\
\end{eqnarray}

and
\begin{eqnarray}
(i<\chi(0)|{\dot \chi}(0)>)^2 = {|<\Psi(0)|{\dot \Psi}(0)>|^2 \over ||\Psi(0)||^4 } + {||{\dot \Psi}(0)|| \over ||\Psi(0)||}^2 - {||{\dot \Psi}(0)|| \over ||\Psi(0)||^3} 2 Re <\Psi(0)|{\dot \Psi}(0)>
\end{eqnarray}
Using  above  expressions  we  can write the constant $ k = [
{ <{\dot \Psi}(0)|{\dot \Psi}(0)> \over ||\Psi(0)||^2 } -
{|<\Psi(0)|{\dot \Psi}(0)>|^2 \over ||\Psi(0)||^4 }]$.

Now,  the next step would be to argue that  the constant $k$ is a
non-negative (whose physical meaning  will  be  discussed).  Note
that  the  transition  probability  of  finding the system in the
initial state has $\tau^2$ dependence,  for  short  times,  which
comes  naturally  from  the  inner product of the vectors defined
over  the  Hilbert space of the quantum system. In order that the
coherent  time  evolution  leads  to   a   survival   probability
approaching   unity   under   repeated   observation   the   {\it
non-negativity of $k$ is crucial}. In the following we provide an
argument based on Hilbert space geometry that it is indeed so.

\par
To  provide  a  physical meaning to the quantity $k$ in a general
situation,  we  take  recourse to some basic ideas of geometry of
quantum  evolution.   Quantum   mechanically,   the   system   is
represented  not just by a vector but by a ray. A ray is a set of
vectors which differ from each other by $U(1)$ phase factors.  If
we  take  a  projection $\Pi: {\cal H} \rightarrow {\cal P}$, then
all  the  points  in  a  ray  project  to  a  single point in the
projective  Hilbert  space  of  the  quantum  system.  Therefore,
geometrically  the  system  is  represented  by  a  point  on the
projective Hilbert space ${\cal P}$ of the  quantum  system.  The
projective  Hilbert  space is formed by taking the set of rays of
the Hilbert space. Let $\{|\Psi> \} \in {\cal H}$, then  consider
the set of non-zero vectors of unit norm $\{|\chi>
=  {|\Psi>  \over  ||\Psi||} \} \in  {\cal  H}^*$  . Thus, the
projective Hilbert space ${\cal P} = {\cal H}^*/U(1)$, where U(1)
is  the group with non-zero complex numbers. The evolution of the
system  is represented by a curve in ${\cal P}$ and if we want to
know how much distance has been travelled  by  the  system  point
$\Pi(|\chi>)$, then we need to know the metric defined on it.

Some  times  back  it  has  been  shown  by Provost and Vallee in
\cite{pv} that the inner product in the Hilbert space can  induce
a  Riemannian  structure on the projective Hilbert of the quantum
system. In studying geometry of quantum evolution  and  geometric
phases the inner product between two state vectors once again plays a
very important role. The  inner  product  or  survival  amplitude
between  two  non-orthogonal states is in general complex number,
which has a magnitude and a phase. The physical  meaning  of  the
magnitude  is  that  its  square  gives us transition probability
(which is of concern here).  The  phase  in  general  contains  a
dynamical  and  a geometric component. The geometric phase is one
of the fundamental discovery in recent times which tells us that  the
wavefunction  of  a  quantum system can acquire a phase depending
solely on the geometry of the  path  in  the  projective  Hilbert
space  of  the  system  \cite{jv}.  It  is  worth mentioning that
quantum Zeno effect does not answer the question: what happens  to
phase  under  repeated measurements? This is an important question
which  was  addressed  only  recently  by  the   present   authors
\cite{as}.  The  quest  of geometric structures led Aharonov and
Anandan \cite{16} to introduce Fubini-Study metric which provides
the total distance  travelled  by  the  system  point  along  the
evolution  curve  in  the  projective  Hilbert  space  of quantum
system. The connection between geometric phases  and  the  metric
structures   were   clarified  by  one  of  the  present  authors
\cite{17}.  With  the  help of the metric structures many quantum
mechanical phenomena has been viewed in a geometric way. Further,
the  Fubini-Study  metric  was  generalised by one of the present
authors  \cite{18}  to  case  of  non-unitary,  non-Schr\"odinger
evolutions  arising  from  the inner product of vectors, which is
given by

\begin{equation}
s^2 = 4\biggl(  1  -
\biggl|\biggl<{\Psi_1   \over  ||\Psi_1||}  \biggl|{\Psi_2  \over
||\Psi_2||}\biggr>\biggr|^2 \biggr)
\end{equation}
Physically, this metric  is a measure of the distance between two
states  in  the  Hilbert space or the corresponding points in the
projective Hilbert space of  the  quantum  system.  It  satisfies
identity,  symmetry,  and  triangle  inequality conditions. It is
invariant under $U(1)$ gauge  transformation  as  well  as  under
generalised  gauge  transformation  (see \cite{18}). When the two
vectors  differ  infinitesimally,  we  obtain  the  infinitesimal
Fubini-Study metric defined as

\begin{equation}
ds^2  =  4\biggl[
\biggl<{d          \over         dt}\biggl({\Psi(t)         \over
||\Psi(t)||}\biggr)\biggl|  {d  \over  dt}\biggl({\Psi(t)   \over
||\Psi(t)||}\biggr)\biggr>    -   \biggl(i\biggl<{\Psi(t)   \over
||\Psi(t)||}\biggr|{d     \over     dt}\biggl({\Psi(t)      \over
||\Psi(t)||}\biggr) \biggr>\biggr)^2 \biggr]dt^2.
\end{equation}

\par
This  generalised  metric  \cite{18}  is valid even when a system
undergoes non-linear,  non-unitary  and  inhomogeneous  evolution
equation.  In  the special case of linear, unitary evolution this
reduces to the Fubini-Study metric defined in \cite{16,17}.  This
metric   is   invariant  under  gauge  transformation  $|\Psi(t)>
\rightarrow e^{i\alpha(t)}  |\Psi(t)>$.  Also,  it  is  invariant
under  generalised  gauge  transformation  $|\Psi(t)> \rightarrow
Z(t) |\Psi(t)>$, with $Z(t)$ being a complex function of non-unit
modulus. Since the modulus of  the  inner  product  is  invariant
under all unitary and anti-unitary transformations, the metric is
so.  Moreover,  this  does not depend on the detailed dynamics of
the system. If we define  the  speed  at  which  the system point
moves on the projective Hilbert space of the  quantum  system  is
$v(t)  =  ds/dt$, then the total distance travelled by the system
point during an arbitrary quantum evolution is given by

\begin{equation}
s  = \int v(t) dt
\end{equation}
where $v(t) =
2[<{\dot \chi}(t)|{\dot \chi}(t)>) - (i<\chi(t)|{\dot
\chi}(t)>)^2 ]^{1 \over 2}$, which can also be written as
$ v = 2[
{ <{\dot \Psi}(t)|{\dot \Psi}(t)> \over ||\Psi(t)||^2 } -
{|<\Psi(t)|{\dot \Psi}(t)>|^2 \over ||\Psi(t)||^4 }]^{1 \over 2}$.
This  geometric  quantity  is  a  reprametrisation  invariant and
depends only on the projected path of the evolving quantum system.

\par
It  follows  that  the  quantity  $k$  appearing  in the survival
probability (5) has a clear physical meaning of being the  square
of  the the speed of transportation of the system point on ${\cal
P}$  evaluated  at  initial  time  i.e.,  $k  =  v^2(0)/4$.   The
non-negativity of the real constant $k$ is therefore guaranted.

\par
With  this  geometrical  interpretation  of  survival probability
$P(\tau)$, if we consider successive measurements in N number  of
steps  which  consists  of  free evolution for a time $\tau$ plus
instantaneous measurements of the observable $O$ at times $\tau_k
= k\tau, k = 1,2,.....N,$ then the  probability  of  finding  the
system in the initial state would be given by

\begin{equation} P(\tau_N) =
\biggl[ P(\tau) \biggr]^N = (1 - \tau^2 v(0)^2/4)^N.
\end{equation}
For  large number of measurements we can approximate the survival
(inhibition)   probability   $P(\tau_N)$   as
\begin{equation}
P(\tau_N) = e^{-\tau^2 v(0)^2 /4N} = e^{- T^2v(0)^2 / 4N}.
\end{equation}
This  shows  that  the  survival  probability  after $N$ steps of
measurement process is a geometric quantity in nature. Therefore,
in the limit $N  \rightarrow  \infty$  the  survival  probability
tends  to  unity. This leads to {\it the QZE for systems governed
by   more   general   dynamical   equations}   (refered   to   as
non-Schr{\"o}dinger)   within   the   collapse  hypothesis.  Only
requirement necessary is that the solutions  of  these  equations
are  square  integrable  functions  and thus they belong to their
respective Hilbert spaces.  This  is  always  the  case  for  the
Schr{\"o}dinger   equation   and   hence   it   seems  that  {\it
Schr{\"o}dinger equation is not a necessary but only a sufficient
requirement for predicting QZE.} Any  other  evolution  equation,
which  could  be  non-linear, non-unitary and inhomogeneous would
also predict QZE as it is related to the geometry of the  Hilbert
space involved.

\par
It  is  interesting  to  investigate the meaning of $v(0)$ in the
special case of linear, unitary Schr{\"o}dinger time evolution of
the quantum system. For Schr{\"o}dinger quantum evolution, with a
time-independent   Hamiltonian   $H$   the    quantity    $<{\dot
\chi}(0)|{\dot    \chi}(0)>$    goes    over    to   $-{1   \over
{\hbar}^2}<\Psi(0)|H^2|\Psi(0)>
= -{1 \over {\hbar}^2}<\Psi(t)|H^2|\Psi(t)>$ and $ i<\chi(0)|{\dot \chi}(0)>$
goes  over  to  ${1  \over  \hbar} <\Psi(0)|H|\Psi(0)> = {1 \over
\hbar} <\Psi(t)|H|\Psi(t)>$. Therefore, in this case the quantity
$v(0)$ is proportional to  the  uncertainty  $\Delta  E$  in  the
energy  of the system, i.e., $v = 2{\Delta E \over \hbar}$.
In other words, for Schr{\"o}dinger time evolution the  speed  of
the  system  point on ${\cal P}$ is decided by the fluctuation in
the energy of the system and one recovers the standard Zeno  type
result   for   survival   probability,   given  by  $P(\tau_N)  =
exp(-{T^2\Delta E^2 \over N})$,\cite{11} which  in  the  ultimate
limit $N \rightarrow \infty$ approaches unity.

\par
Next, we give a criterion for the QZE to occur with finite number
of  measurements  (since  we  are  not  taking  an ideal limit $N
\rightarrow \infty$). For a finite number $N$ of measurements, it
follows from Eq.(11) that
\begin{equation}
dP(\tau_n)/dN > 0.
\end{equation}
This  may  be  taken as a criterion for the occurance of QZE. The
geometrical  nature of the survival probability $P(\tau_N)$ along
with  the condition (13) implies that for the QZE to occur, it is
necessary that the dynamics of a quantum system is  characterised
by  an  evolution equation whose solutions spans a Hilbert space.
In essence the  evolution  of  a  quantum  system  can  be  quite
arbitrary  between  two  successive  measurements and QZE follows
purely from the collapse of the wavefunction and the geometry  of
the Hilbert space.

\par
Using  these  ideas we illustrate how can one observe the quantum
Zeno effect for system described by  non-linear  and  non-unitary
equations.   As  a  first  example  we  consider  the  non-linear
Schr{\"o}dinger equation

\begin{equation}
i\hbar{\partial \Psi({\bf x}, t)\over \partial t}  =  -{\hbar^2  \over  2m}
\nabla^2 \Psi({\bf x},t) + V({\bf x}) \Psi({\bf x},t) - b|\Psi|^2 \Psi({\bf x},t)
\end{equation}
where  $V({\bf  x})$  is  the  confining potential and $b$ is the
non-linearity parameter whose  meaning  depends  on  the  problem
considered.  This  equation has been used in various contexts. It
describes the dynamics of  quasi-particles  in  condensed  matter
physics.  More  recently, this equation has been of much interest
in understanding the behaviour of  Bose-Einstein  condensates  in
the confining trap potential \cite{19}.

\par
This  equation preserves the norm of the wavefunction, that is, $
||{\dot \Psi}|| = 0$ for all times. We will show  that  a  system
described by Eq(14) conforms to the short time behaviour as given
by  Eq.(5).  As  a particular case, we consider a one-dimensional
form of Eq.(14) with $V(x) = 0$, written in the form

\begin{equation}
i\hbar{\partial \Psi(x, t)\over \partial t}  +
{\hbar^2  \over 2m}{\partial^2 \Psi(x, t)\over \partial x^2} + b|\Psi|^2 \Psi(x,t) = 0
\end{equation}
where we assume  that  $b > 0$.
This equation admits a propagating soliton-like solution given by
\begin{equation}
\Psi(x,  t)  =  ({a  \over  b})^{1 \over 2} \eta  e^{-i(\omega  t  - ux/a)}
sech[\eta(x - ut)]
\end{equation}
where  $\eta  =  {1  \over  a}(u^2 - 2a\omega)^{1 \over 2}$, $a =
{\hbar^2 \over m}$. It is easy to check that $||\Psi||^2 = (2\eta
a /b)$ which shows that the norm of  the  wavefunction  does  not
change with time. Using Eq.(11) we can see that the speed at which
the system point move on the projective Hilbert space is given by
$v^2  =  4(\eta u)^2 / 3$. The short time survival probability is
given by $P(\tau) = (1 - \tau^2 \eta^2 u^2 / 3)$. Therefore,  the
survival probability after N steps of measurement is $P(\tau_N) =
exp(  -  \eta^2 u^2 T^2 /3N )$ which satisfies (13) and hence the
QZE occurs. It is interesting to note that  the  Zeno  effect  in
this  case  would  imply  an inhibition of the propagation of the
solitonic wave.

\par
A  second  example  is  provided  by  a  model  proposed by Gisin
\cite{20} for description of quantum dissipative  systems.   This
phenomenological non-linear equation reads as
\begin{equation}
i\hbar{\partial |\Psi(t)>\over \partial t}  =
H|\Psi(t)>  + i\lambda \biggl({<\Psi|H|\Psi> \over <\Psi|\Psi>} -
H \biggr)|\Psi(t)>,
\end{equation}
where $H$ is the Hamiltonian of the system and $\lambda \ge 0$ is
a  dimensionless  damping  constant. This equation also preserves
the norm of  the  state  vector  during  its  evolution.  Another
interesting  feature  is  that  this equation retains most of the
conventional interpretation of the quantum theory.

\par
We apply Gisin's equation to a two-level system. Let us  consider
a  simple  case  of  the  model  of  a two-level atom interacting
continuously with a coherent field. The Hamiltonian of the system
is given by
\begin{equation}
H = \hbar \omega S_+ S_- - {\hbar \alpha \over 2}(S_+ e^{-i\omega
t} + S_-e^{i\omega t}),
\end{equation}
where $S_+ = |g><e|,  S_- = |e><g|;  |g>, |e>$ being the ``ground''
and ``excited''  states  of  the  atom  and  $\alpha$,  the  Rabi
frequency  related  to  the  amplitude  of the driving field. For
simplicity, we ignore the spontaneous emission from  the  excited
state  and  also  assume  that the field is in resonance with the
atomic transition frequency $\omega$. The general  state  of  the
system belongs to a Hilbert space of dimension two and is given by
\begin{equation}
|\Psi(t)> = a(t)|g>_+ b(t) e^{-i\omega
t}|e>
\end{equation}
and inserting (18) and (19) in Gisin's equation (17), we obtain
\begin{eqnarray}
{\dot   a(t)} & = &  {\alpha   \over   2}(\lambda   +   i)b(t)   +
                     \lambda[\omega|b(t)|^2 - {\alpha   \over   2}(a^*(t)b(t) + a(t) b^*(t))]a(t) \nonumber\\
{\dot   b(t)} & = &  {\alpha   \over   2}(\lambda   +   i)a(t)   +
                     \lambda[\omega(|b(t)|^2 - 1)- {\alpha   \over   2}(a^*(t)b(t) + a(t) b^*(t))]b(t).
\end{eqnarray}

\par
It is easy to obtain the short-time behaviour of the solutions of
the above  equations.  Thus,  if  we  assume  that  the  atom  is
initially   in   the   ground  state  $|g>$,  then  the  survival
probability $P(\tau)$ is given by
\begin{equation}
P(\tau)  = 1 - {\alpha^2 \over 4}(\lambda^2 + 1)\tau^2.
\end{equation}

The  system  point  moves  on the projective Hilbert space (which
is a sphere $S^2$ for two-level system)
of the quantum system with a speed $v = \alpha (\lambda^2 +  1)^{1/2}$
Also, we can calculate the
Fubini-Study metric during the time $T$ which is given by $s =
\alpha (\lambda^2 +  1)^{1/2}T$
Therefore, with the increase of number of  measurements  the
survival probability approaches $
P(\tau_N)   =  exp( - {\alpha^2 \over 4}(\lambda^2 + 1)T^2 / N) =
exp(-s^2/4N)$,
and the quantum Zeno effect
in this case implies that the atom remains in the ground state.
It may be interesting to note that the presence  of  the  damping
constant  $\lambda$ in  the  survival  probability  enhances  the
quantum Zeno
effect.

\par
In  conclusion,  we  have shown in this paper that the short time
quadratic  behaviour  of  the transition probability is a natural
consequence  of  the  inner  product  of the Hilbert space, using
which the former is defined. The survival probability is  related
to  the  speed  of  transportaion  of  the  system  point  on the
projective Hilbert space of the quantum  system. It is shown that
the  Schr{\"o}dinger  time  evolution  between   two   successive
measurements  is  not a necessary but only a sufficient condition
to predict quantum Zeno effect. Any non-linear,  non-unitary  and
inhomogeneous  evolution equation could also predict quantum Zeno
effect  within  the  von  Neumann's  collapse   postulate.   More
importantly, it seems that the presence of non-linear and damping
parameters  in  the  system  can  enhance the Zeno process with a
finite  number  of  measurement  pulses.  It  would   indeed   be
interesting  to  study  the  role  of  non-linearity  and damping
parameters in real quantum Zeno type experiments in future.

\vskip 3cm

{\centerline {\bf ACKNOWLEDGEMENTS}}

The authors wish to thank Prof.  E.  C.  G.  Sudarshan  for  some
stimulating  discussions on the subject matter of Zeno effect for
non-linear quantum systems.\\

\renewcommand{\baselinestretch}{1}

\end{document}